 \documentclass[10pt,preprint]{aastex}  %compact preprint style (MJA)

\def\ang{\AA}
\def\arcsec{\hbox{$^{\prime\prime}$}}

\def\gapprox{\lower.4ex\hbox{$\;\buildrel >\over{\scriptstyle\sim}\;$}}
\def\lapprox{\lower.4ex\hbox{$\;\buildrel <\over{\scriptstyle\sim}\;$}}

\def\ref#1{\par\noindent\hangindent1cm {#1}}

\shortauthors{ASCHWANDEN AND SANDMAN 2010}
\shorttitle{Coronal Magnetid Field Bootstrapping. I.}

\begin{document}
%{\sl  Manuscript, 2010-Apr-23}

\title{ Bootstrapping the Coronal Magnetic Field with STEREO: 
	I. Unipolar Potential Field Modeling	}

\author{Markus J. Aschwanden$^{1,2}$ and Anne W. Sandman$^{1,2}$}

\affil{$^1)$ Solar and Astrophysics Laboratory,
	Lockheed Martin Advanced Technology Center, 
        Dept. ADBS, Bldg.252, 3251 Hanover St., Palo Alto, CA 94304, USA; 
        (e-mail: aschwanden@lmsal.com)}

\affil{$^2)$ Physics and Astronomy Department,
	Rice University,
	Houston, TX, USA.}

\date{(Received ... April 2010; accepted ...}

\begin{abstract}
We investigate the recently quantified misalignment of $\alpha_{mis}
\approx 20^\circ-40^\circ$ between the 3-D geometry of
stereoscopically triangulated coronal loops observed with STEREO/EUVI
(in four active regions)
and theoretical (potential or nonlinear force-free) magnetic field models 
extrapolated from photospheric magnetograms. We develop an efficient method
of bootstrapping the coronal magnetic field by forward-fitting a 
parameterized potential field model to the STEREO-observed loops.
The potential field model consists of a number of unipolar magnetic 
charges that are parameterized by decomposing a photospheric magnetogram
from MDI. The forward-fitting method yields a best-fit magnetic field model 
with a reduced misalignment of $\alpha_{PF} \approx 13^\circ-20^\circ$.
We evaluate also stereoscopic measurement errors and find a contribution
of $\alpha_{SE}\approx 7^\circ-12^\circ$, which constrains the residual
misalignment to $\alpha_{NP}=\alpha_{PF}-\alpha_{SE}\approx 5^\circ
-9^\circ$, which is likely due to the nonpotentiality of the active regions.
The residual misalignment angle $\alpha_{NP}$ of the potential field due 
to nonpotentiality is found to correlate with the soft X-ray flux of 
the active region, which implies a relationship between electric currents 
and plasma heating.
\end{abstract}


\section{       INTRODUCTION 					}

The STEREO mission provides us an unprecedented view
of the solar corona, enabling us for the first time to fully constrain
the three-dimensional (3-D) geometry of the coronal magnetic field. 
Stereoscopic triangulation
of coronal loops has been conducted at small STEREO spacecraft separation 
angles ($\alpha_{sep} \lapprox 10^\circ$), for several active
regions observed with STEREO A(head) and B(ehind) in April and May 2007
(Aschwanden et al.~2008a,b; 2009). The reconstructed 3-D geometry of
STEREO-observed coronal loops has been compared with theoretical
magnetic field models based on extrapolations from photospheric 
magnetograms, using
nonlinear force-free field (NLFFF) models (DeRosa et al.~2009), as well as
potential and stretched potential field models (Sandman et al. 2009), but 
surprisingly it turned out that the two types of magnetic field lines
exhibited an average misalignment angle of $\alpha_{mis} \approx
20^{\circ}-40^{\circ}$, regardless of what type of theoretical magnetic
field model was used. From this dilemma it was concluded that a more realistic
physical model is needed to quantify the transition from the non-force-free
photospheric boundary condition to the nearly force-free field at the base
of the solar corona (DeRosa et al.~2009). 

At this junction, it is not clear what a viable method is to obtain a 
force-free boundary of the magnetic field at the coronal base, or how 
to correct the non-force-free magnetograms. However, the stereoscopic 
triangulation supposedly provides the correct 3-D directions of the 
magnetic field ${\bf B}({\bf x})$, which together with Maxwell's 
equation of divergence-freeness ($\nabla {\bf B}=0)$, constrain also 
the absolute values of the field strengths. 
In this Paper I we choose a magnetic field model that is
defined in terms of multiple unipolar charges. An approach in terms
of multiple dipoles is employed in Paper II (Sandman and Aschwanden 2010).
Since both unipolar or dipolar magnetic fields represent potential 
magnetic fields that fulfill the divergence-free condition, the 
superposition of multiple unipolar and dipolar magnetic field
components fulfill the same condition. We develop a numerical 
code of such a parameterized divergence-free magnetic field that can be 
forward-fitted to the 3-D geometry of stereoscopically triangulated 
coronal loops. So, the simple goal of this study is to evaluate how
closely the stereoscopically observed loops can be modeled in terms
of potential fields, a goal that was already attempted with Skylab
observations (Sakurai and Uchida 1977).
Modeling with non-potential fields, such as
nonlinear force-free field (NLFFF) models, will be considered in 
future studies. 

This paper is organized as follows: The definition of a parameterized
potential field is described in Section 2, the development and tests
of a numeric magnetic field code and the results of forward-fitting 
to stereoscopically triangulated loops is presented in Section 3, 
and conclusions follow in Section 4. An alternative approach with 
dipolar magnetic fields is the subject of Paper II (Sandman and 
Aschwanden 2010). 

\section{       THEORY AND DEFINITIONS 				}

\subsection{	Divergence-Free Magnetic Field			}

Since the coronal plasma-$\beta$ parameter is generally less than unity
(Gary 2001), the magnetic pressure exceeds the thermal pressure, 
and thus all soft X-ray or EUV-emitting plasma that fills or flows 
through coronal flux tubes traces out the coronal field. 
Consequently, stereoscopic triangulation of EUV loops
provides the correct 3-D field directions along coronal loops. We denote
the normalized 3-D field direction along a loop with the unity vector
${\bf b}(s)$, which is parameterized as a function of a loop length 
coordinate $s = s({\bf x})=s(x,y,z)$, starting at the footpoint position 
$s_0=s(x,y,0)$ at the base of the corona, 
\begin{equation}
         {\bf b}(s) = {{\bf B}(s) \over B(s)}
         = {[B_x, B_y, B_z]
         \over \sqrt{B_x^2+B_y^2+B_z^2} } \ ,
\end{equation}
where the magnetic field is defined by the cartesian components
${\bf B}(s) = [B_x(s), B_y(s), B_z(s)]$.
However, the absolute magnitude of the magnetic field strength, 
$B(s) = |{\bf B}(s)|$, is not known a priori.
For a physical solution of the magnetic field, Maxwell's equation 
of a divergence-free magnetic field has to be satisfied,
\begin{equation}
                \nabla {\bf B} = \left( {\partial B_x \over dx} +
                {\partial B_y \over dy} + {\partial B_z \over dz} \right)
                = 0 \ ,
\end{equation}
which (in its integral form) corresponds to the theorem of magnetic flux
conservation along a fluxtube,
\begin{equation}
                \Phi(s) = \int B(s) dA = const \ ,
\end{equation}
where $A(s)=\int dA$ is the integral over the cross-sectional area of a
fluxtube defined in perpendicular direction to the magnetic field line
(at loop position $s$). Therefore, since the stereoscopic triangulation 
defines the magnetic field directions in adjacent flux tubes, it defines 
also the divergence of the field and the relative change of the magnetic
field strength $B(s)$ along the flux tubes, and this way implicitly defines 
also the isogauss surfaces perpendicular to each fluxtube, and therefore
the full 3-D vector field ${\bf B}$, except for a scaling constant. 
The derivation of a 3-D magnetic field ${\bf B}$ in an active region,
\begin{equation}
        {\bf B}({\bf x}) = B({\bf x}) {\bf b}({\bf x}) \ ,
\end{equation}
requires also the knowledge of the scalar function $B({\bf x})$ in every 
3-D location ${\bf x}$, which we constrain with a forward-fitting method
of a divergene-free field model.

A divergence-free 3-D magnetic field model ${\bf B}({\bf x})$
can be parameterized by a superposition of divergence-free fields,
because the divergence-free condition is linear (or Abelian), i.e., if two
components ${\bf A}$ and ${\bf B}$ fulfill $\nabla {\bf A}=0$ and
$\nabla {\bf B}=0$, then also their sum is divergence-free,
$\nabla ({\bf A}+{\bf B}) = \nabla {\bf A} + \nabla {\bf B} = 0$.
Divergence-free magnetic field components are, for instance,
a parallel field, unipolar fields (a magnetic charge with
spherical isogauss surfaces and a field that falls off with the
square of the distance), dipole fields, quadrupolar fields, other
multi-pole representations, or any potential field. The Abelian 
property warrants that any superposition of divergence-free fields 
is also divergence-free. Specifically, we will use divergence-free
potential field models that consist of either
{\it i)} multiple unipolar charges (in this Paper I), 
or {\it ii)} multiple dipoles (in Paper II). 

Our philosophy is the following: We will employ magnetic field models 
of the category of potential-field models, which are divergence-free
by definition. We use particular potential field models that can be
quantified with a finite number of free parameters. Potential field models 
are not as general as non-potential and force-free field (NLFFF) models.
However, since both potential and NLLLF models currently exhibit an equally
poor misalignment with observed EUV loops, we need first to investigate
whether the misalignement between observations and {\sl any} theoretical 
model can be mimized for the simplest class of magnetic field models,
such as potential field models. If our approach proves to be successfull, 
refinements with non-potential or NLLLF models can then be pursued in 
the future along the same avenue (e.g., see Conlon and Gallagher 2010).

\subsection{	Multiple Unipolar Magnetic Charges 	  }

Unipolar potential fields often provide a good approximation to
the magnetic field of sunspots, and thus can be used also for an
active region that is composed of a finite number of spot-like magnetic
polarities. Conceptually, a unipolar field can be considered as an 
approximation to the upper half of a vertically positioned dipole.
The simplest representation of a unipolar magnetic field that is
a potential field, and hence fulfills Maxwell's divergence-free condition,
is a spherically symmetric field that drops off with an $r^{-2}$-dependence
with distance, so it has a potential function that drops off 
with $r^{-1}$,
\begin{equation}
	\Phi(r)= - \Phi_0 \left( {z_0 \over r} \right) \ ,
\end{equation}
where $\Phi_0$ is the potential field value at the solar surface
vertically above the buried magnetic charge in depth $z_0 < 0$. 
Of course, the extrapolated magnetic field is only computed in the
coronal domain $z > 0$, so that no ``magnetic monopole'' exists in
the solar corona. The magnetic field model of a single magnetic charge
requires 4 parameters: the maximum value of the potential field $\Phi_0$
and the location $(x_0, y_0, z_0)$ of the buried charge, having a
distance $r$ to any point $(x,y,z)$ in the solar corona, 
\begin{equation}
         r   = \left[(x-x_0)^2+(y-y_0)^2+(z-z_0)^2 \right]^{1/2} \ . 
\end{equation}
The resulting magnetic field has then only a radial component $B_r$
in direction of ${\bf r}$,
\begin{equation}
	B_r (r) =\nabla \Phi(r) = B_0 \left({z_0 \over r}\right)^2 \ ,
\end{equation}
with the surface field strength $B_0=\Phi_0/z_0$. This unipolar potential
field fulfills the divergence-free condition, as it can be calculated
from the Laplacian operator of the potential function,
\begin{equation}
	\nabla {\bf B}(r) =
	\Delta \Phi (r) = 
	{1 \over r^2} {\partial \over \partial r}
	\left( r^2 {\partial \Phi(r) \over \partial r} \right) = 0 \ .
\end{equation}
The fulfillment of the divergence-free condition can also be verified
from the conservation of the magnetic flux theorem (Eq.~3), if the
envelope of a fluxtube is defined by radial field lines, so that the
cross-sectional area $A(s) \propto B(s)^{-2}$ remains constant for $s=r$. 
In our first model we employ a superposition of $N$ multiple unipolar
charges,
\begin{equation}
        {\bf B}({\bf x}) = \sum_{j=1}^N {\bf B}_j({\bf x})
        = \sum_{j=1}^N  B_j 
	\left({z_j \over r_j}\right)^2 {{\bf r}_j \over r_j} \ ,
\end{equation}
in terms of the vector ${\bf r}_j = [(x-x_j), (y-y_j), (z-z_j)]$. 
For a single unipolar charge, the field lines will all be straight
lines in radial direction away from the buried charge, which can
approximate open-field regions. Burying multiple magnetic charges
of opposite magnetic polarity, however, can mimic closed-field
regions. An example is given in Fig.~1, where we compare the
magnetic field of a dipole with that of a combination of two
unipolar charges with opposite magnetic polarity. Actually,
the two magnetic field models become identical when the two
unipolar charges are moved close together at the location of the dipole
moment, as it can be shown mathematically. Although the two models are 
equivalent in the far-field approximation, a combination of two unipolar
charges (with $2 \times 4 = 8$ free parameters) allows more general
solutions than a single dipole (with 6 free parameters), especially
in the case of strongly asymmetric fields (sunspots) or open-field
regions, as they exist is most active regions.  

\subsection{	Definition of Misalignment Angle			}

The forward-fitting of our analytical magnetic field model 
to a set of observed magnetic field vectors ${\bf b}={\bf B}/B$ (e.g.,
using stereoscopically triangulated loop coordinates),  
is the task of optimizing the free parameters of the analytical
model until the best match with the observed field lines is obtained.
For the evaluation of the goodness or consistency of the analytical magnetic
field models ${\bf B}^{theo}$ with the observed field line model 
${\bf B}^{obs}$,
we define the 3-D misalignment angle $\alpha_{mis}$,
which is defined by the scalar product between the two field vectors
${\bf B}^{theo}({\bf x})$ and ${\bf B}^{obs}({\bf x})$,
\begin{equation}
        \alpha_{mis}({\bf x}) =
        cos^{-1} \left({ {\bf B}^{theo}({\bf x}) \cdot
        {\bf B}^{obs}({\bf x}) \over
        |{\bf B}^{theo}({\bf x})|\ |{\bf B}^{obs}({\bf x})| }\right) \ ,
\end{equation}
or equivalently, between the unity field vectors 
${\bf b}^{theo}({\bf x})$ and ${\bf b}^{obs}({\bf x})$,
\begin{equation}
        \alpha_{mis}({\bf x}) =
        cos^{-1} \left({ {\bf b}^{theo}({\bf x}) \cdot
        {\bf b}^{obs}({\bf x}) \over
        |{\bf b}^{theo}({\bf x})|\ |{\bf b}^{obs}({\bf x})| }\right) \ .
\end{equation}

This parameter is a single value at every spatial point ${\bf x}$,
which can be averaged at the coronal base or over the lengths of the
observed field lines, at $n_p$ positions, 
\begin{equation}
        <\alpha_{mis}> = \left[ {1 \over n_p} \sum_{i=1}^{n_p}
         \alpha_{mis}^2(x_i, y_i, z_i) \right]^{1/2} \ ,
\end{equation}
which is similar to a $\chi^2$-criterion. Since a unipolar magnetic
charge can be parameterized with 4 free parameters ($x_j, y_j, z_j, B_j)$
(Eq.~9), a model with $n_p$ components has $n_p=4 n_c$ free parameters. 

\section{       	NUMERICAL CODE AND RESULTS 			}

Our strategy to bootstrap the coronal magnetic field with stereoscopic data
consists of the following steps: (1) we create a model of subphotospheric
magnetic charges by deconvolving an observed photospheric magnetogram 
into point charges, which defines the unit vectors of 
our parameterized theoretical magnetic field model ${\bf B}^{theo}(x,y,0)$;
(2) we perform stereoscopic triangulation for a
set of coronal loops observed with STEREO/EUVI, which are quantified
in terms of directional field vectors ${\bf b}^{obs}$; 
(3) we forward-fit the theoretical magnetic field model ${\bf b}^{theo}$
with a number of free parameters in the setup of unipolar magnetic charges 
by minimizing the mean misalignment angle 
$\Delta \alpha_{mis}({\bf b}^{theo}, {\bf b}^{obs})$, and
which yields a best-fit solution ${\bf B}^{theo}$.  We can then compare 
the minimized misalignment of the bootstrapped best-fit model 
with those of standard methods based on extrapolation of the 
the photospheric boundaries using a magnetogram, e.g., with the {\sl Potential
Field Source Surface (PFSS)} model. The procedure is illustrated in Fig.~2,
where a dipolar EUV loop is observed and a magnetic field model is constructed
from two unipolar charges. By adjusting the field strength of
the second unipolar charge from $B_2=0.1$ to $B_2=0.08$, the misalignment
angle between the observed EUV loop and the model field can be reduced
from $\alpha_{mis}=20^0$ (Fig.~2 top) to $\alpha_{mis}=0^0$ (Fig.~2,
bottom). Note that the adjusted field is still a potential field and 
divergence-free, but represents a better match to the observed EUV loop.

\subsection{       Data Selection 			 	}

We select four active regions observed with STEREO/EUVI and the 
{\sl Michelson Doppler Imager (MDI)} (Scherrer et al.~1995) 
onboard the {\sl Solar and
Heliospheric Observatory (SoHO)}: 2007 April 30, May 9, May 19, and Dec 11.
The first AR 10953 (2007 Apr 30) is identical to the case previously
analyzed with STEREO and Hinode (DeRosa et al.~2009; Sandman et al.~2009).
The second AR 10955 (2007 May 9) was subject of the first stereoscopically
triangulated coronal loops, temperature and density measurements, and
stereoscopic tomographic reconstruction (Aschwanden et al.~2008a,b, 2009;
Sandman et al.~2009). The third AR 10953 (2007 May 9) displayed a small flare 
(during 12:40-13:20 UT) as well as a partial filament eruption during the 
time of observations, and was featured in a few studies (Li et al.~2008; 
Liewer et al.~2009; Sandman et al.~2009). The fourth AR 10978
(2007 Dec 11) is also subject of recent magnetic field modeling (Aad
Van Ballegooijen and Alec Engell, private communication 2010).  
Some details of these four active regions are given in Table 1, such as
the heliographic position of the AR center, the magnetic area for fluxes 
of $B>100$ G, the minimum and maximum field strengths, and the total 
unsigned magnetic flux.

\subsection{       Parameterization of Magnetic Field Model 	 	}

All four active regions were observed with the 
{\sl Michelson Doppler Imager (MDI)} onboard 
the {\sl Solar and Heliospheric Observatory (SoHO)}, which provides full-disk 
MDI magnetograms with a pixel size of 2\arcsec. Subimages encompassing the
active region of interest are shown in Fig.~3 (left column), 
with quadratic field-of-view 
sizes ranging from 145 to 339 pixels, or 0.3 to 0.7 solar radii, respectively.
In order to create a realistic 3D magnetic field model we 
decompose the partial magnetograms into a number of $n_c=200$ positive and 
negative Gaussian 2-D components, using an iterative 2-D Gaussian fitting 
scheme that determines local maxima in decreasing order of field strengths. 
The composite magnetogram of these 200 decomposed sources is shown in Fig.~2 
(middle column), and the difference to the original magnetogram is also shown
(right column). For each of the 200 magnetic source components we store the
peak magnetic field value $B_i$, the center position ($x_j, y_j$), and the 
half widths $w_i$ at full maximum. For a parameterization in terms of unipolar
charges, however, we need to convert the half width $w_i$ into the corresponding
depth $z_j$ at which the unipolar charge is buried. From the definition of the full 
width at half maximum (FWHM) at $|x-x_j|=w_j$ of a unipolar field (Eq.~9),
we have for the vertical magnetic field component $B_z=B_r \cos{\vartheta}
=B_r (z_j/r_j)$ (see geometry in Fig.~4),
\begin{equation}
	B_z(x_j+w_j)  
	=B_r(x_j+w_j) \cos{\vartheta} 
	=B_0 \left({z_j \over r}\right)^3
	=B_0 \left({z_j^2 \over w_j^2 + z_j^2}\right)^{3/2}
	={B_0 \over 2} \ ,
\end{equation}
with $\vartheta$ the angle between the vertical and a surface ring with
radius $w_j$ (Fig.~4), from which we can calculate the dipole depth $z_j$,
\begin{equation}
	z_j = - {w_j \over \sqrt{2^{2/3}-1}} \approx - 1.30 \ w_j \ ,
\end{equation}
which corresponds approximately to the half width $w_j$ of the fitted Gaussian
component. Since we have now all input parameters $(x_j, y_j, z_j, B_j)$,
$j=1, ... ,n_c$, for a definition of a magnetic field model with multiple
unipolar magnetic charges (Eq.~9), we can calculate the full 3D magnetic
field by superimposing the fields of all components. A particular field line 
is simply calculated by starting with the field at a footpoint position 
and by iterative stepping in the field direction, until the field line
returns to the solar surface (in case of closed field lines) or to a 
selected boundary of the computation box (for open field lines).

\subsection{	Stereoscopic Triangulation of EUV Loops			}

For each of the four active regions we triangulate as many loops as can be
discerned with a highpass filter in image pairs from STEREO/EUVI A and B.
The method of stereoscopic triangulation is described in detail in
Aschwanden (2008a), from which we use identical loop coordinates for
AR 10955 observed on 2007 May 9. The stereoscopic triangulation requires
accurate coalignment of both EUVI/A and B images in an epipolar coordinate
system. Furthermore, we transform the 3-D loop coordinates measured in
the epipolar coordinate system with the line-of-sight of the EUVI/A spacecraft
into the coordinate system of MDI, which sees the Sun from the Langrangian
point L1, almost in the same direction as seen from Earth. The advantage of
transforming the EUVI loop coordinates into the MDI reference frame is 
the direct modeling of the longitudinal magnetic field component $B_{\parallel}$
in the MDI reference frame, without requiring any knowledge of the absolute
magnetic field strength $B$, which is model-dependent, i.e., it depends on
the choice of the extrapolation method (potential, force-free, or NLFFF) from
photospheric magnetograms (which measures only the longitudinal component
$B_{\parallel}$). Some parameters of the stereoscopic triangulation procedure
are listed in Table 1, such as the heliographic coordinates of the active
region, the spacecraft separation angle, and the number of triangulated EUV
loops (varying between 70 and 200 per active region, combined from all three
coronal wavelengths, 171, 195, and 284 \ang ). 3-D representations of the
stereoscopically triangualted EUV loops are shown in Figs.~5-8 (with blue
color), seen along the line-of-sight of MDI (in greyscale maps of Figs.~5-8)
and in the two orthogonal directions (sideview and topview in Figs.~5-8).
The height range of stereosopically triangulated loops generally does not
exceed 0.1 solar radii, due to the drop of dynamic range in flux for 
altitudes in excess of one hydrostatic scale height.

\subsection{	Forward-Fitting of Potential Field Models		}

After we have parameterized the magnetic field with $n_c$ unipolar charges,
each one defined by 4 parameters $(x_j, y_j, z_j, B_j)$, using the procedure
of iterative decomposition of a photospheric magnetogram as described in
Section 3.2, we can vary these free model parameters to adjust it to the
3D geometry of the stereoscopically triangulated loops. However, since we
typically represent an MDI magnetogram with $n_c \approx 200$ components,
we have $n_p=4 n_c \approx 800$ free parameters, which are too many to
optimize independently. Standard optimization procedures, such as the Powell
algorithm (Press et al.~1986), allow for optimization of $\approx 10-20$
free parameters, with good convergence behavior if the initial guesses are
suitably chosen. Therefore we choose 10-20 small circular zones (with typical
radii of $r_{sep} \approx 0.01-0.02$ solar radii) containing the strongest 
field regions in the magnetogram of equal magnetic polarity and optimize
the magnetic field strengths $B_j^{opt}=q_j B_j$ with a common correction 
factor $q_j$ in each zone. This is an empirical optimization of the lower
boundary of the coronal magnetic field, optimized by varying the (decomposed)
photospheric MDI magnetogram in such a way that the resulting potential field
more closely matches the stereoscopically triangulated loops. The results of the
best fits, based on the mimization of the median misalignment angle 
(Eqs.~11-12), are shown in Figs.~5-9, in form of (red) model field lines that
are extrapolated at the same locations as the footpoints of the stereoscopic
loops (blue in Figs.~5-9). The distribution of all misalignment angles,
evaluated at about 80 locations of every loop is shown in form of histograms
at the bottom of Figs.~5-9. Each histogram is characterized with a gaussian
peak, which are found to have a mean value and standard deviation of
$\alpha=14.3\pm11.5$ (Fig.5: 2007 Apr 30), 
$\alpha=13.3\pm 9.3$ (Fig.6: 2007 May 9), 
$\alpha=20.3\pm16.5$ (Fig.7: 2007 May 19), and 
$\alpha=15.2\pm12.3$ (Fig.8: 2007 Dec 11). 
If we compare these misalignment angles in the range of 
$\alpha_{mis}\approx 13^\circ-20^\circ$ 
with the previously measured values using a potential field source surface (PFSS)
model ($\alpha_{mis}\approx 19^\circ-36^\circ$; Sandman et al.~2009), or using
a nonlinear force-free field (NLFFF) model 
($\alpha_{mis}\approx 24^\circ-44^\circ$; DeRosa et al.~2009), we see an improvement 
of the misalignment angle by about a factor of two. 
This is a remarkable result that demonstrates that the magnetic field at the
lower boundary of the corona can be bootstrapped with stereoscopic measurements
and with a suitable parameterization of a potential field model. Although the 
extrapolation from a photospheric magnetogram should be unique, if there is 
no data noise present, an infinite number of potential field solutions can be 
obtained depending on how the boundary field is parameterized (e.g., by
unipolar charges or dipolar magnetic moments). In our case, 
for every variation of the $n_p\approx 800$ free parameters, a slightly 
different field with a different misalignment to the stereosocpic loops 
is obtained. The PFSS code is designed to compute the potential field of 
the entire (front and backside) Sun, and thus has a relatively coarse 
spatial resolution of typically $\approx 1^\circ$ (one heliographic degree,
i.e., 12 Mm) for standard computations, but the small-scale magnetic field
should not matter too much for our large-scale loops ($\gapprox 10$ Mm).
Hence, the best-fit potential-field bootstraps a force-free photospheric 
boundary field that is significantly different from the observed 
photospheric magnetograms which is observed in a non-force-free
zone. 

\subsection{	Convergence Behavior 				}

The remaining misalignment between our best-fit potential field solution and
the stereoscopic loops could be due to three sources of errors: (1) The
bootstrapping method did not converge to the best solution; (2) The 3D coordinates
of the stereoscopically triangulated loops have some error; or (3) The real coronal
magnetic field that is represented by the stereoscopic loops could be non-potential.
We consider the convergence of our code as satisfactory, because we ran many attempts
for each case with different initial conditions and obtained about the same minimum
misalignment. One possibility to vary the initial conditions is to vary the number
of (gaussian) unipolar components. Fig.~9 shows the convergence behavior as a
function of the number of (decomposed) unipolar components, where we computed a best-fit
potential field solution for $n_c=10, 20, 50, 100, 200, 500$ unipolar components
for each of the 4 active regions. Convergence to the minimum misalignment value 
typically recquires $n_c \approx 10$ components for the simplest dipolar active regions
(2007 May 9 or Dec 11), and $n_c=50-100$ for more complex active regions 
(2007 Apr 30 or May 19). 

\subsection{	Estimates of Stereoscopy Error			}

We investigate the second possible source of errors that could contribute to
the measured misalignment, i.e., errors associated with the stereoscopic triangulation
method. The errors of stereoscopic triangulation have been discussed in Aschwanden
et al.~(2008a), which depend on (1) the ratio of the stereoscopic parallax angle to
the spatial resolution of the instrument, and thus on the spacecraft separation angle,
(2) the angle between the loop segment and East-West direction in the epipolar plane,
being largest for loop segments parallel to the epipolar plane, and (3) the proper 
identification of a corresponding loop in image B to a selected loop in image A.
While the first two sources of errors can be formally calculated, the correspondence
problem is difficult to quantify. Since stereoscopic triangulation cannot be
accomplished in an automated way at present time, the error of identifying corresponding
loops could be estimated from the scatter obtained with different observers, but this
is time-consuming. Here we pursue another approach that is based on a self-consistency 
test. The procedure works as follows. If a sufficient large number of loops are
triangulated in an active region, there should be for every triangulated loop (which
we call primary loop) a neighbored (secondary) loop with an almost parallel 
direction. Whatever the direction of the true magnetic field in the same
neighborhood is, both the primary $(i)$ and the secondary STEREO loop $(j)$ should have 
a similar misalignment angle with the local magnetic field, 
$\Delta \alpha_i \approx \Delta \alpha_j$, or 
$|\Delta \alpha_i - \Delta \alpha_j| \approx 0$, in order to be self-consistent.
Therefore we can define a stereoscopic error (SE) angle $\Delta \alpha_{SE}$ by
averaging these differences in misalignements over all STEREO loop positions
with suitable weighting factors $w_{ij}$,
\begin{equation}
	\Delta \alpha_{SE} = { \sum_{i,j} |\alpha_i^{mis}-\alpha_j^{mis}| w_{ij}
		             \over \sum_{i,j} w_{ij} } \ ,
\end{equation} 
where the index $i$ runs over all loop positions for each primary loop and
the index $j$ runs over all loop positions of secondary loops, excluding
the primary loop. For the weighting factor $w_{i,j}$ we should give less weight
to more distant neighbors, because the true magnetic field and thus the
local misalignment angle is likely to increase with distance. Thus we should
choose a negative power of the relative distance $d_{ij}$,
\begin{equation}
	w_{ij} = {1 \over d_{ij}^p} = 
		 [(x_i-x_j)^2+(y_i-y_j)^2+(z_i-z_j)^2]^{-p/2} \ ,
\end{equation}
where $p$ is a power index. If the index $p$ is small, say $p=1$ the weighting is
reciprocal to the distance and the long-range neighbors are relatively strongly
weighted. If the index $p$ is large, say $p=10$, the short-range neighbors are
relatively strongly weighted while the longe-range neighbors have almost no weight.
So we expect that the so-defined stereoscopic error monotonously decreases with
short-range weighting towards higher power indices $p>1$. In Fig.~10 we show
the stereoscopic error calculated with Eqs.~(15-16) as a function of the power
index from $p=1$ to $p=20$, for all 4 active regions. Indeed, the stereoscopic
error monotonously decreases from $p=1$ to $p=12$, because we give progressively
less weight to the long-range misalignments. However, from $p=12$ towards $p=20$
the error increases again, probably because of the inhomogeneity of the closest
neighbors at the shortest range and the excessive weighting to the very nearest
neighbors. However, the plateau in the range of $p\approx 5-15$ is a good indication
that we measure a stable value at the minumum of $p\approx 12$. From this we
obtain a misalignment contribution of $\Delta \alpha_{SE}=9.4^\circ$
(2007 Apr 30), $5.7^\circ$ (2007 May 9), $11.5^\circ$ (2007 May 19), and
$7.2^\circ$ (2007 Dec 11), as listed in Table 2. 

\subsection{	Nonpotentiality of Magnetic Field 			}

A diagram of the various misalignment angles is shown in Fig.~11, which we
plot as a function of the maximum GOES soft X-ray flux during the observing 
interval. The GOES flux was dominanted by the emission from the active 
regions analyzed here, since there was no other comparable active region 
present on the solar disk during the times of the observations. 
The case of 2007 May 9 with the lowest GOES shows a single dipolar 
structure, while the case of 2007 May 19 with the highest GOES flux exhibits 
multiple dipolar groups. 
The misalignment angles $\alpha_{PFSS}$ indicate those obtained with
the {\sl potential field source surface (PFSS)} code, $\alpha_{NLFF}$ those
with the {\sl nonlinear force-free field (NLFFF)} code (only for 2007 Apr 30), 
$\alpha_{PFU}$ those optimized with the {\sl potential field with unipolar 
charge (PFU)} code, and $\Delta \alpha_{SE}$ indicates the stereosocpic errors.
As a working hypothesis we might attribute the remaining residuals $\alpha_{NP}$ 
to the non-potentiality of the active regions,  
\begin{equation}
	\alpha_{NP} = \alpha_{PFU} - \Delta \alpha_{SE} \ ,
\end{equation}
for which we measure the values:
$\alpha_{NP}=4.9$ (2007 Apr 30), 
$\alpha_{NP}=5.7$ (2007 May 9), 
$\alpha_{NP}=8.8$ (2007 May 19), and
$\alpha_{NP}=7.2$ (2007 Dec 11), listed also in Table 2.
We consider these values as a new method to quantify the nonpotentialiy
of active regions. We find that quiescent active regions that contain simple 
dipoles (2007 Apr 30 and May 9) have small misalignments of $\alpha_{NP} 
\approx 5^\circ-6^\circ$ with best-fit potential field models,
while more complex active regions (2007 May 19 and Dec 11) have somewhat larger
non-potential misalignments in the order of 
$\approx 7^\circ-9^\circ$ compared with best-fit potential field models.

While we express thge degree of nonpotentiality in terms of a mean
misalignment angle here, other measures are the ratios of the total
nonpotential to the potential energy in an active region, which amounts
up to $E_{NP}/E_P \lapprox 1.32$ in one flaring active region 
(Schrijver et al.~2008).

\subsection{	Quiescent and Flaring Active Regions		}

Since the four investigated active regions have quite different misalignment
angles, and also the inferred nonpotentiality of the magnetic field varies
significantly, we quantify the activity level of the active regions from
the soft X-ray flux measured by GOES. Fig.~12 shows the GOES light curves 
for the 4 active regions during the times of stereoscopic triangulation. 
The lowest GOES flux is measured for the AR of 2007 May 9, with a level of
$10^{-7.6}$ W m$^{-2}$ (GOES class A4), while the highest GOES flux is measured
for the AR of 2007 May 19, which has a flare occurring during the observing period
with a GOES flux of $10^{-6.0}$ W m$^{-2}$ (GOES class C0). The three active
regions with the low soft X-ray flux levels appear to be quiescent, judging
from the GOES flux profile, or may be subject to micro-flaring at a low level.

There is a clear correlation between the soft X-ray level of the active region 
(when it was on disk) and the overall misalignment 
angle (Fig.~11), as well as with the misalignment angle $\alpha_{NP}$
attributed to the non-potentiality, varying from $\alpha_{NP} \approx 5^\circ$
for the lowest GOES A-class levels to $\alpha_{NP} \approx 9^\circ$ for
an active region with a GOES C-class flare. A higher soft X-ray flux generally
means a higher heating rate with stronger impulsive heating or flaring.
A higher degree of non-potentiality, on the other hand, indicates the presence
of a higher level of electric currents (which are non-potential). Therefore,
the observed correlation suggests a physical relationship between the electric
currents in an active region and the amount of heating input. This is not
surprising, since evidence for current-carrying emerging flux was demonstrated
previously for H$\alpha$ and soft X-ray structures that are non-potential 
(e.g., Leka et al.~1996; Jiao et al.~1997; Schmieder et al.~1996).
Our measurement of the degree of non-potentiality
with the magnetic field misalignment averaged over the entire active region,
is a very coarse technique, but a more detailed investigation of the misalignment
in separate parts of the active region that are quiescent or flaring will be
pursued in Paper II. 

\section{		Conclusions				}

The agreement between theoretical magnetic field models of active regions
in the solar corona with the true 3-D magnetic field as delineated from
the stereoscopic triangulation of coronal loops in EUV wavelengths has
never been quantified until the recent advent of the STEREO mission.
To everybody's surprise, the average misalignment between the theoretical
and observed magnetic field was quite substantial, in the amount of
$\alpha_{mis} \approx 20^\circ-40^\circ$ for both potential and nonlinear
force-free field models (DeRosa et al.~2009; Sandman et al.~2009). 
In this study we investigate the various contributions of this large 
misalignment for four different active regions observed with STEREO
and arrive at the following conclusions:

\begin{enumerate}
\item{The amount of misalignment can be reduced to about half of the values
	for potential-field models optimized by a bootstrapping method that
	minimizes the field directions with the stereoscopically triangulated
	loops. Our potential-field model is parameterized with $\approx 200$ 
	unipolar charges per active region, whose positions and field strengths
	are approximately derived from a gaussian decomposition of a photospheric 
	magnetogram, and then varied until a best fit is obtained. The best-fit
	potential field model has an improved misalignment of $\alpha_{PFU} 
	\approx 13^\circ-20^\circ$. Because the best-fit potential field model
	defines an improved magnetic field boundary condition at the bottom of
	the corona, the difference to the observed photospheric magnetogram 
	contains information on the currents between the photosphere and the
	base of the force-free corona.}
\item{We estimate the misalignment contriubtion caused by stereoscopic correlation
	errors from self-consistency measurements between the magnetic field misalignments
	of adjacent loops. We find contributions in the order of 
	$\Delta \alpha_{SE} \approx 7^\circ-12^\circ$.}
\item{We estimate the contributions to the field misalignment due to 
	non-potentiality caused by electric currents from the residuals between
	the best-fit potential field and the stereoscopic triangulation errors
	and find misalignment contributions in the order of
	$\alpha_{NP}\approx 5^\circ-9^\circ$.}
\item{The overall average misalignemnt angle between potential field models
	and stereoscopic loop directions, as well as the contribution to the
	misalignment due to non-potentiality, are found to correlate with the
	soft X-ray flux of the active region, which suggests a correlation
	between the amount of electric currents and the amount of energy
	dissipation in form of plasma heating in an active region.}
\end{enumerate}

In this study we identify for the first time the contributions to the misalignment
of the magnetic field, in terms of optimized potential field models, non-potentiality
due to electric currents, and stereoscopic triangulation errors. These results open
up a number of new avenues to improve theoretical modeling of the coronal magnetic
field. First of all, optimized potential field models can be found that represent
a suitable lower boundary condition at the base of the force-free corona, which 
provides a less computing-expensive method than nonlinear force-free codes. 
Second, methods can be developed that allow us to localize electric currents in the 
non-force-free photophere and chromosphere. Third, the misalignment angle can be
used as a sensitive parameter to probe the evolution of current dissipation,
energy build-up in form of non-potential magnetic energy in different 
quiescenct and flaring zones of active regions. The high-resolution magnetic field
data from Hinode and {\sl Solar Dynamics Observatory} provide excellent opportunities
to obtain better theoretical models of the coronal magnetic field using our
bootstrapping method, which is not restricted to stereoscopic data only,
but can also be applied to single-spacecraft observations. 
	
\medskip
Acknowledgements: We are grateful to helpful discussions with Marc DeRosa
and Allen Gary. This work was partially supported by the NASA contract NAS5-38099
of the TRACE mission and by NASA STEREO under NRL contract N00173-02-C-2035.
The STEREO/SECCHI data used here are produced by an international consortium of
NRL, LMSAL, RAL, MPI, ISAS, and NASA. The MDI/SoHO data were produced by
the MDI Team at Stanford University and NASA.

\section*{REFERENCES} 	%References

\def\ref#1{\par\noindent\hangindent1cm {#1}}
\def\aap {{\sl A\&A}\ } % Astronomy and Astrophysics
\def\apj {{\sl Astrophys. J.}\ } %The Astrophysical Journal
\def\sp  {{\sl Solar Phys.}\ } % Solar Physics
\def\ssr {{\sl Space Science Rev.}\ } % Space Science Reviews 

\small
\ref{Aschwanden, M.J., 2004, {\sl Physics of the Solar Corona - 
	An Introduction}, Praxis Publishing Ltd., Chichester UK, 
	and Springer, Berlin.}
\ref{Aschwanden, M.J., Wuelser, J.P., Nitta, N., and Lemen, J.R.:
        2008a, \apj {\bf 679}, 827.}
\ref{Aschwanden, M.J., Nitta, N.V., Wuelser, J.P., and Lemen, J.R.:
        2008b, \apj {\bf 680}, 1477.}
\ref{Aschwanden, M.J., Nitta, N.V., Wuelser, J.P., Lemen, J.R.,
        and Sandman, A.: 2009, \apj {\bf 695}, 12.}
\ref{Conlon, P.A. and Gallagher, P.T. 2010, ApJ (submitted).}
\ref{DeRosa M.L., Schrijver,C.J., Barnes,G., Leka,K.D., Lites,B.W., Aschwanden,M.J.,
        Amari,T., Canou,A., McTiernan,J.M., Regnier,S., Thalmann,J., Valori,G., 
        Wheatland,M.S., Wiegelmann,T., Cheung,M.C.M., Conlon,P.A., Fuhrmann,M., 
        Inhester,B., and Tadesse,T. 2009, \apj {\bf 696}, 1780.}
\ref{Gary, A. 2001, Solar Phys. 203, 71.}
\ref{Jiao, L., McClymont, A.N., and Mikic, Z. 1997, \sp 174, 311.}
\ref{Leka, K.D., Canfield, R.C., McClymont, A.N., and Van Driel-Gesztelyi, L.
	1996, \apj 462, 547.}
\ref{Li, Y., Lynch, B.J., Stenborg, G., Luhmann, J.G., Huttunen, K.E.J., Welsch, B.T.,
        Liewer, P.C., and Vourlidas, A. 2008, \apj {\sl 681}, L37-L40.}
\ref{Liewer, P.C., DeJong, E.M., Hall, J.R., Howard, R.A., Thompson, W.T., Culhane, J.L.,
        Bone,L., and VanDriel-Gesztelyi,L. 2009, \sp {\sl 256}, 57-72.}
\ref{Press, W.H., Flannery, B.P., Teukolsky, S.A., and Vetterling, W.T. 1986,
 	{\sl Numerical recipes. The Art of Scientific Computing},
 	Cambridge University Press: New York.}
\ref{Sakurai, T. and Uchida, Y. 1977, \sp {\bf 52}, 397. }
\ref{Sandman, A.W., Aschwanden, M.J., DeRosa, M.L., Wuelser, J.P.,
	and Alexander, D. 2009, \sp {\bf 259}, 1.}
\ref{Sandman, A.W. and Aschwanden, M.J., 2010, (in preparation),
	(Paper II).}
\ref{Scherrer, P.H. et al. 1995, Solar Phys. 162, 129.}
\ref{Schmieder, B., Demoulin, P., Aulanier, G., and Golub,L. 1996, 
	\apj 467, 881.}
\ref{Schrijver,C.J., M.L. DeRosa, T. Metcalf, G. Barnes, B. Lites, T. Tarbell, 
	J. McTiernan, G. Valori, T. Wiegelmann, M.S. Wheatland, T. Amari, 
	G. Aulanier, P. Demoulin, M. Fuhrmann, K. Kusano, S. Regnier, and 
	J.K. Thalmann 2008, \apj 675, 1637.}

\clearpage

%%%%%%%%%%%%%%%%%%%%%%%%%%%% TABLE %%%%%%%%%%%%%%%%%%%%%%%%%%%%%%%%%%%

\begin{deluxetable}{lllrrrrr}
%\rotate
%\tabletypesize{\normalsize}
%\setlength{\tablecolsep}{1pt}
\tabletypesize{\footnotesize}
\tablecaption{Data selection of four Active Regions observed with STEREO/EUVI
and SOHO/MDI.}
\tablewidth{0pt}
\tablehead{
\colhead{Active}&
\colhead{Observing}&
\colhead{Observing}&
\colhead{Spacecraft}&
\colhead{Number}&
\colhead{Magnetic}&
\colhead{Magnetic}&
\colhead{Magnetic}\\
\colhead{Region}&
\colhead{date}&
\colhead{times}&
\colhead{separation}&
\colhead{of EUVI}&
\colhead{area (B$>$100 G)}&
\colhead{field strength}&
\colhead{flux}\\
\colhead{Active}&
\colhead{}&
\colhead{(UT)}&
\colhead{angle (deg)}&
\colhead{(loops)}&
\colhead{($10^{20}$ cm$^2$)}&
\colhead{B(G)}&
\colhead{($10^{22}$ Mx)}}
\startdata
10953 (S05E20)	&2007-Apr-30 	&23:00-23:20	&5.966	&200	&24.2	&[-3134,+1425]& 8.7\\
10955 (S09E24)	&2007-May-9	&20:30-20:50	&7.129	&70	&4.4	&[-2396,+1926]& 1.6\\
10953 (N03W03)	&2007-May-19	&12:40-13:00	&8.554	&100	&12.2	&[-2056,+2307]& 4.0\\
10978 (S09E06)	&2007-Dec-11	&16:30-16:50	&42.698	&87	&8.2	&[-2270,+2037]& 4.8\\
\enddata
\end{deluxetable}

%%%%%%%%%%%%%%%%%%%%%%%%%%%%%%%%%% TABLE 2 %%%%%%%%%%%%%%%%%%%%%%%%%%%%%%%%%%%%%%%

\begin{deluxetable}{lllll}
%\rotate
%\tabletypesize{\normalsize}
%\setlength{\tablecolsep}{1pt}
\tablecaption{Misalignment statistics of four analyzed active regions.}
\tablewidth{0pt}
\tablehead{
\colhead{Parameter}&
\colhead{2007-Apr-30}&
\colhead{2007-May-9}&
\colhead{2007-May-19}&
\colhead{2007-Dec-11}}
\startdata
Misalignment NLFFF$^1$ 	&24$-$44	&	 	&	       & 	      \\ 
Misalignment PFSS$^2$  	&25  $\pm$ 8	&19  $\pm$6 	&36  $\pm$13   &32  $\pm$10   \\ 
Misalignment PFU$^3$ 	&14.3$\pm$11.5	&13.3$\pm$9.3 	&20.3$\pm$16.5 &15.2$\pm$12.3 \\ 
Median       PFU$^3$ 	&20.0     	&16.2 	        &25.8          &15.7          \\ 
Stereoscopy error$^4$ 	& 9.4     	& 7.6 	        &11.5          & 8.9          \\ 
Nonpotentiality$^5$ 	& 4.9     	& 5.7 	        & 8.8          & 7.2          \\ 
GOES soft X-ray flux$^6$&$10^{-7.3}$    &$10^{-7.6}$ 	&$10^{-6.0}$   &$10^{-6.9}$   \\
GOES class              &A7             &A4 		&C0            &B1	      \\
\enddata
\par\noindent $^1)$ Measured with nonlinear force-free field code (DeRosa et al.~2009),
\par\noindent $^2)$ Measured with potential field source surface code (Sandman et al.~2009),
\par\noindent $^3)$ Measured with unpolar potential field model (this study),
\par\noindent $^4)$ Measured from inconsistency between adjacent loops,
\par\noindent $^5)$ Difference of mean misalignment in unipolar model$^2)$,
		stereoscopy error$^4)$,  
\par\noindent $^6)$ Goes flux in units of [W m$^{-2}$].
\end{deluxetable}

\clearpage

%%%%%%%%%%%%%%%%%%%%%%%%%%%% FIGURES %%%%%%%%%%%%%%%%%%%%%%%%%%%%%%%%%%%

\begin{figure} 
\plotone{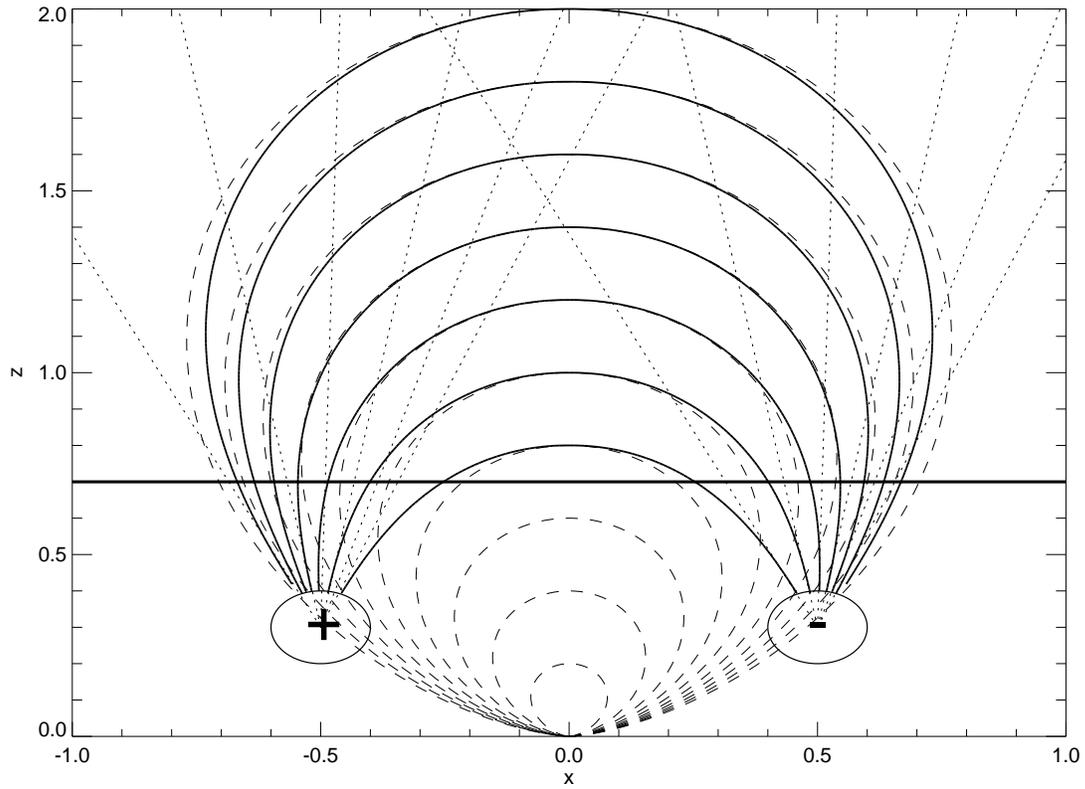}
\caption{The magnetic field of a symmetric dipole (dashed lines)
is shown, together with the field resulting from the superimposition 
of two unipolar magnetic charges (solid lines). The two field models
become identical once the two unipolar charges are moved to the
location of the dipole moment at position $(x,y)=(0,0)$. The radial
field of each unipolar positive charge is also shown for comparison
(dotted lines).}
\end{figure}

\begin{figure} 
\plotone{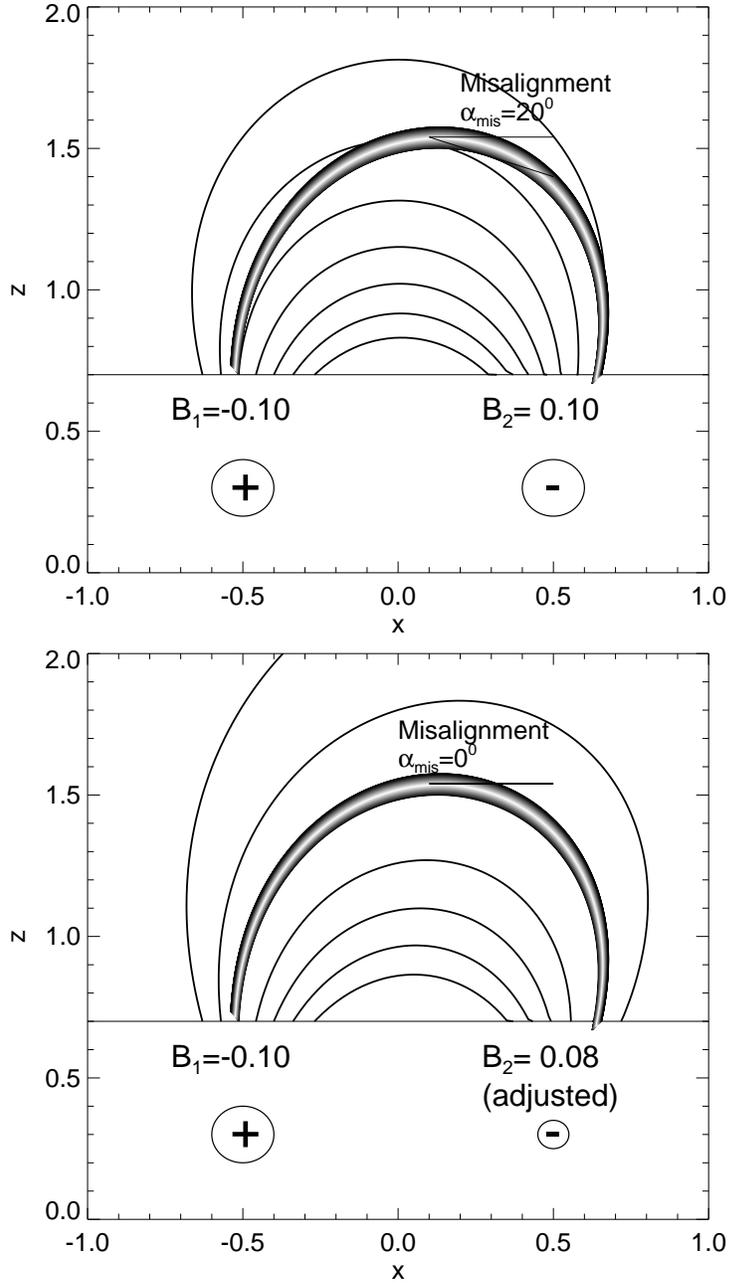}
\caption{{\sl Top:} A potential dipole field is calculated with two
unipolar magnetic charges buried in equal depth and equal magnetic
field strength, but opposite polarization ($B_1=-0.1$, $B_2=+0.1$).
An asymmetric EUV loop is observed at the same location (grey torus)
with a misalighment of $\alpha_{mis}=20^0$ at the top of the loop.
{\sl Bottom:} The magnetic field strength of the right-hand side unipolar
charge is adjusted (to $B_2=0.08$) until the misalignment of the loop
reaches a minimum of $\alpha_{mis}=0^0$.}
\end{figure}

\begin{figure} 
\plotone{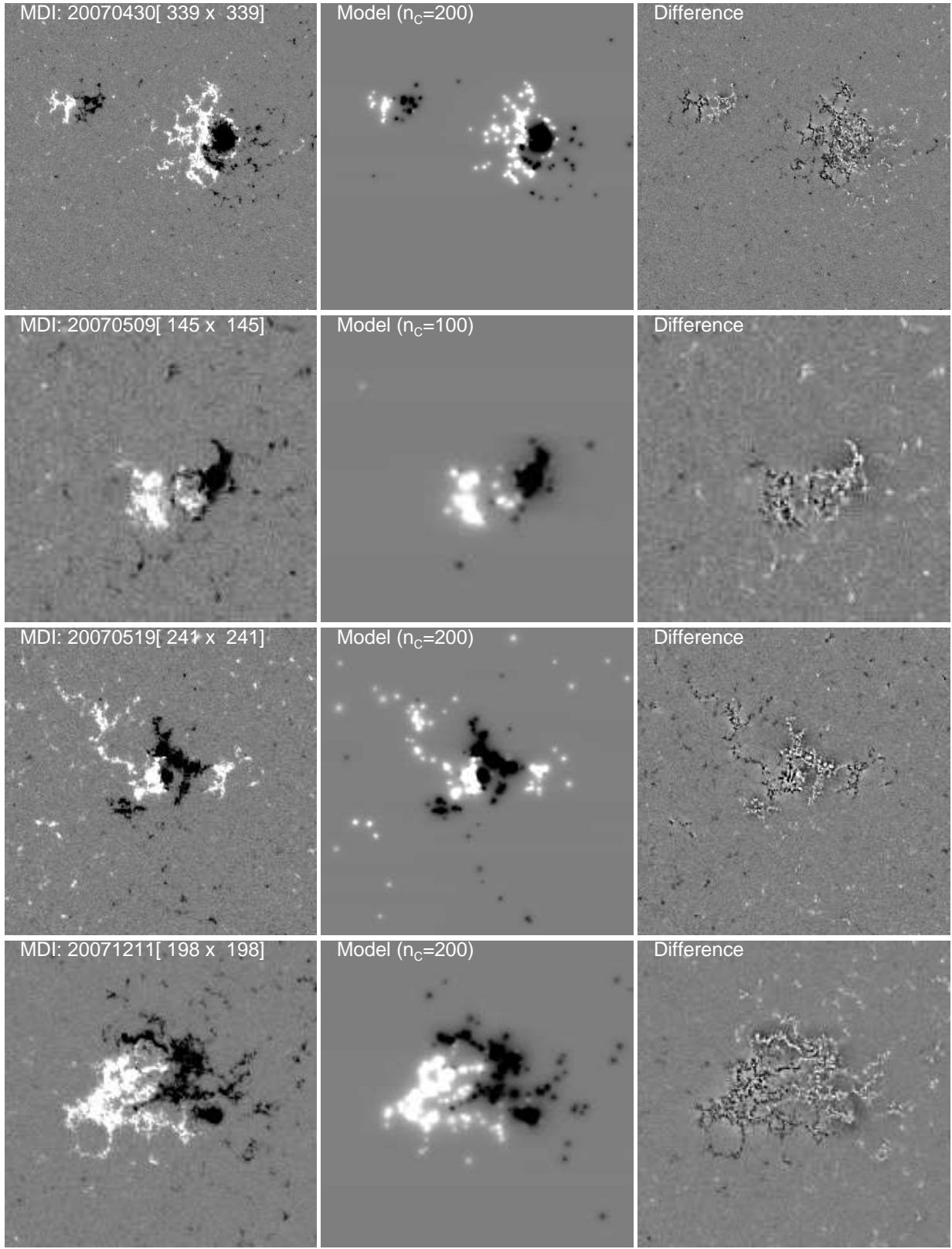}
\caption{{\sl Left column:} Photospheric magnetograms observed with 
SOHO/MDI on four different dates. {\sl Middle:} The magnetic
field is decomposed into typically $n_C=200$ unipolar charges and the
model displays the longitudinal magnetic field strength on the same
grey scale. {\sl Right columne:} Difference between observed MDI
magnetogram and model.}
\end{figure}

\begin{figure} 
\plotone{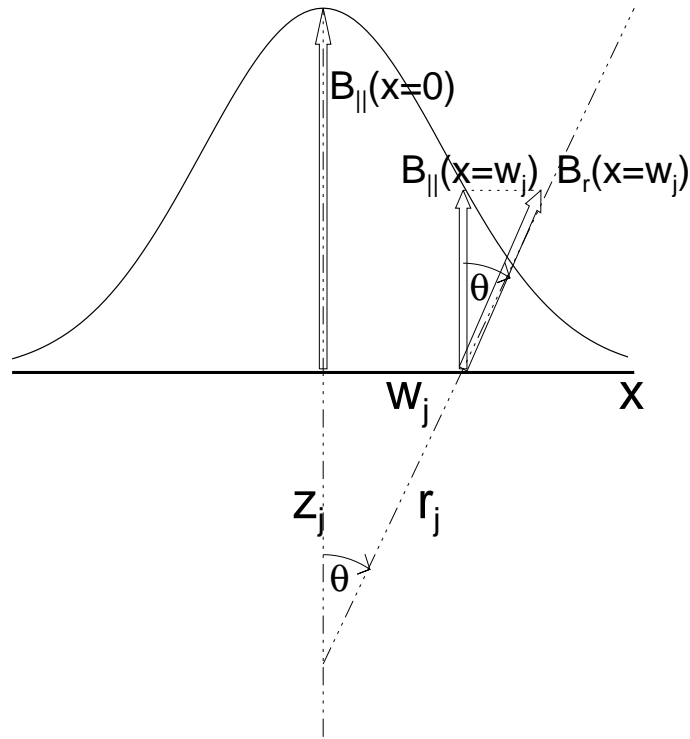}
\caption{Cross-section of gaussian magnetic field component showing the
geometric relation between the half width $w_j$ and depth $z_j$ of a unipolar
charge. The radial magnetic field $B_r$ drops off quadratically with the
distance $r_j$, while the vertical component $B_z = B_r 
\cos{\vartheta}$ is forshortened by a factor $\cos{\vartheta}=z_j/r_j$.}
\end{figure}

\begin{figure} 
\plotone{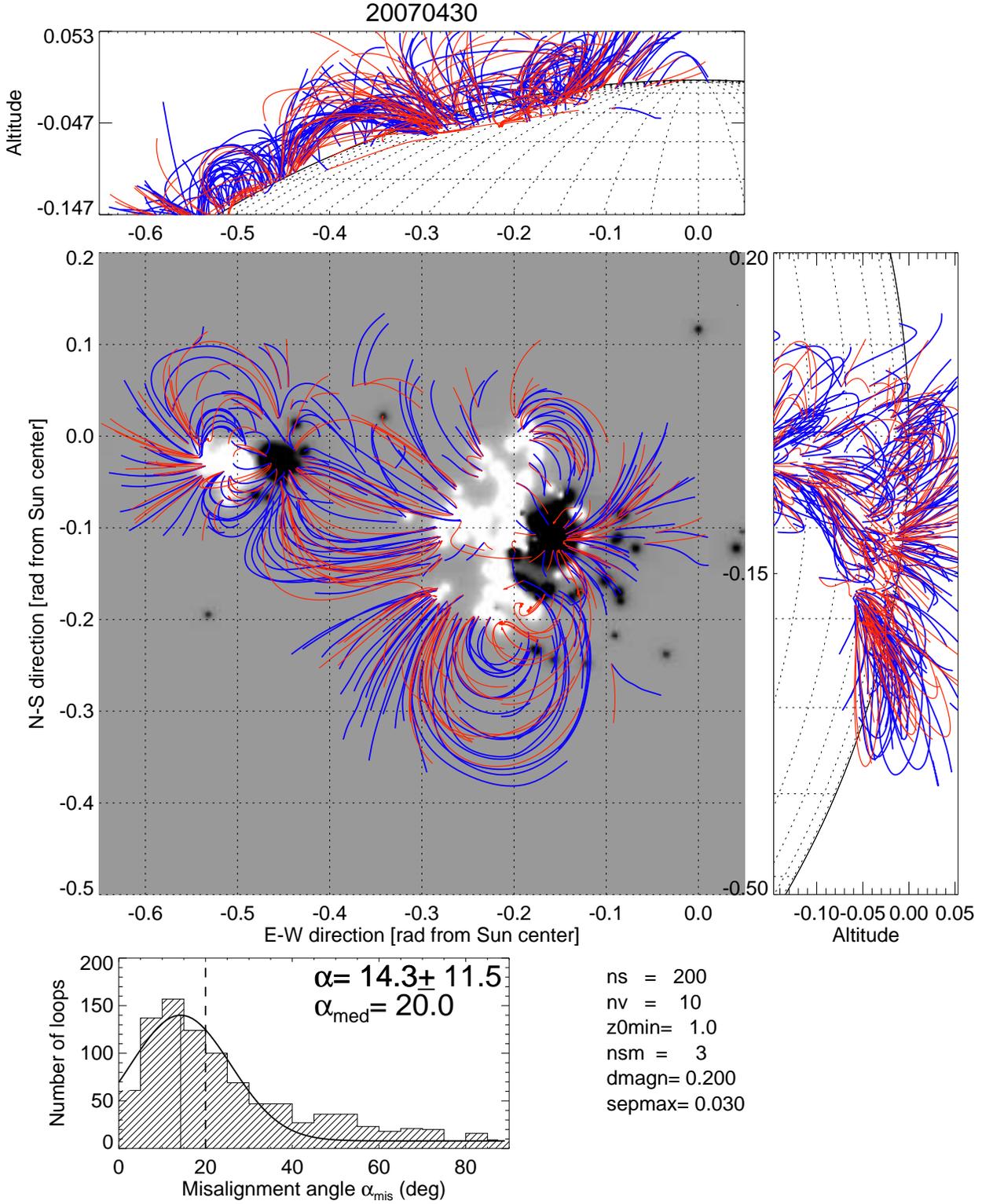}
\caption{Best-fit potential field model of AR observed on 2007 Apr 30.
The stereoscopically triangulated loops are shown in blue color, while
field lines starting at identical footpoints as the STEREO loop 
extrapolated with the best-fit potential field (composed of $n_c=200$
unipolar magnetic charges) are shown in red. Side views are shown in
the top and right panels. A histogram of misalignment angles measured
between the two sets of field lines is displayed in the bottom panel.}
\end{figure}

\begin{figure} 
\plotone{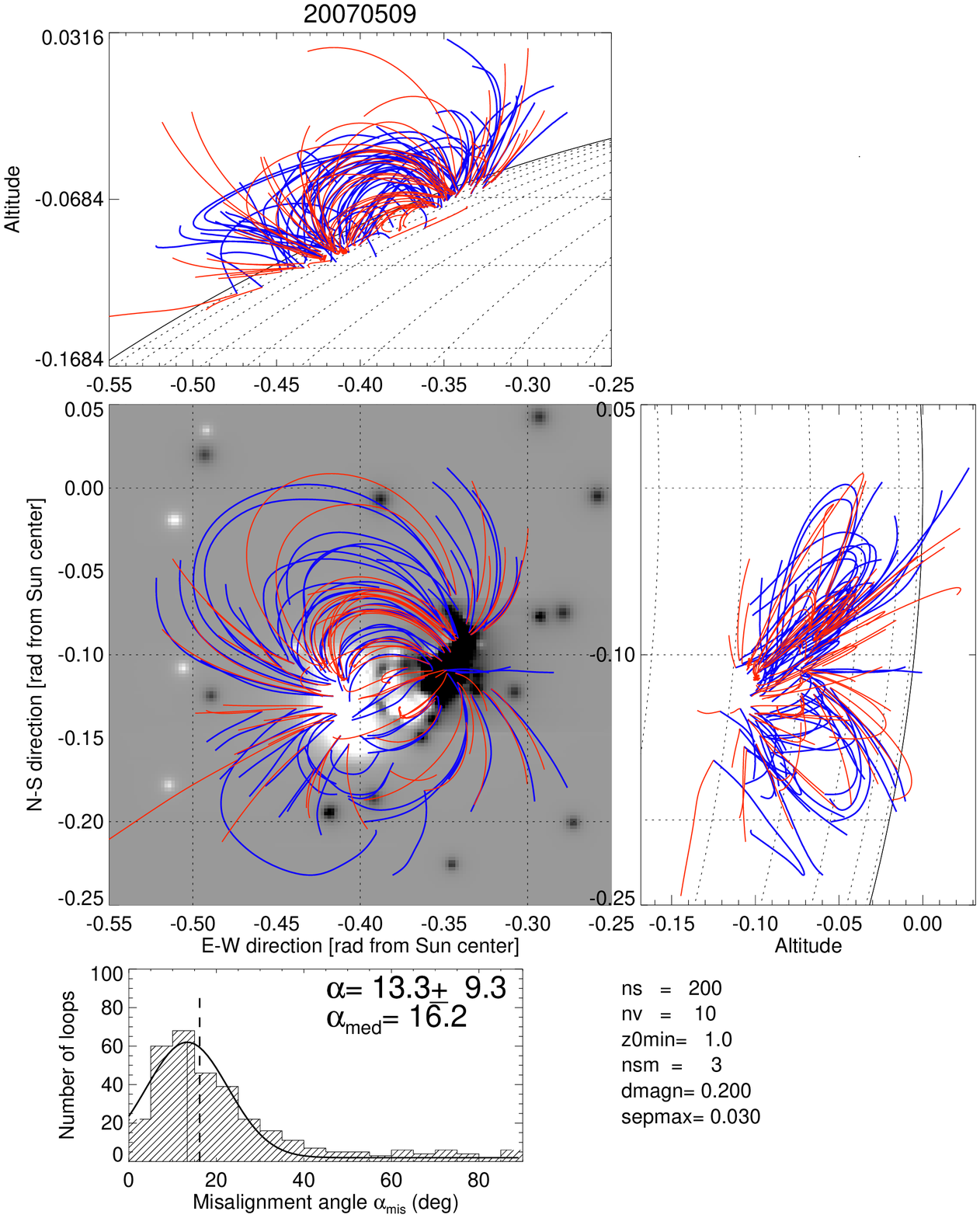}
\caption{Best-fit potential field model of AR observed on 2007 May 9.
Otherwise similar representation as Fig.~5.}
\end{figure}

\begin{figure} 
\plotone{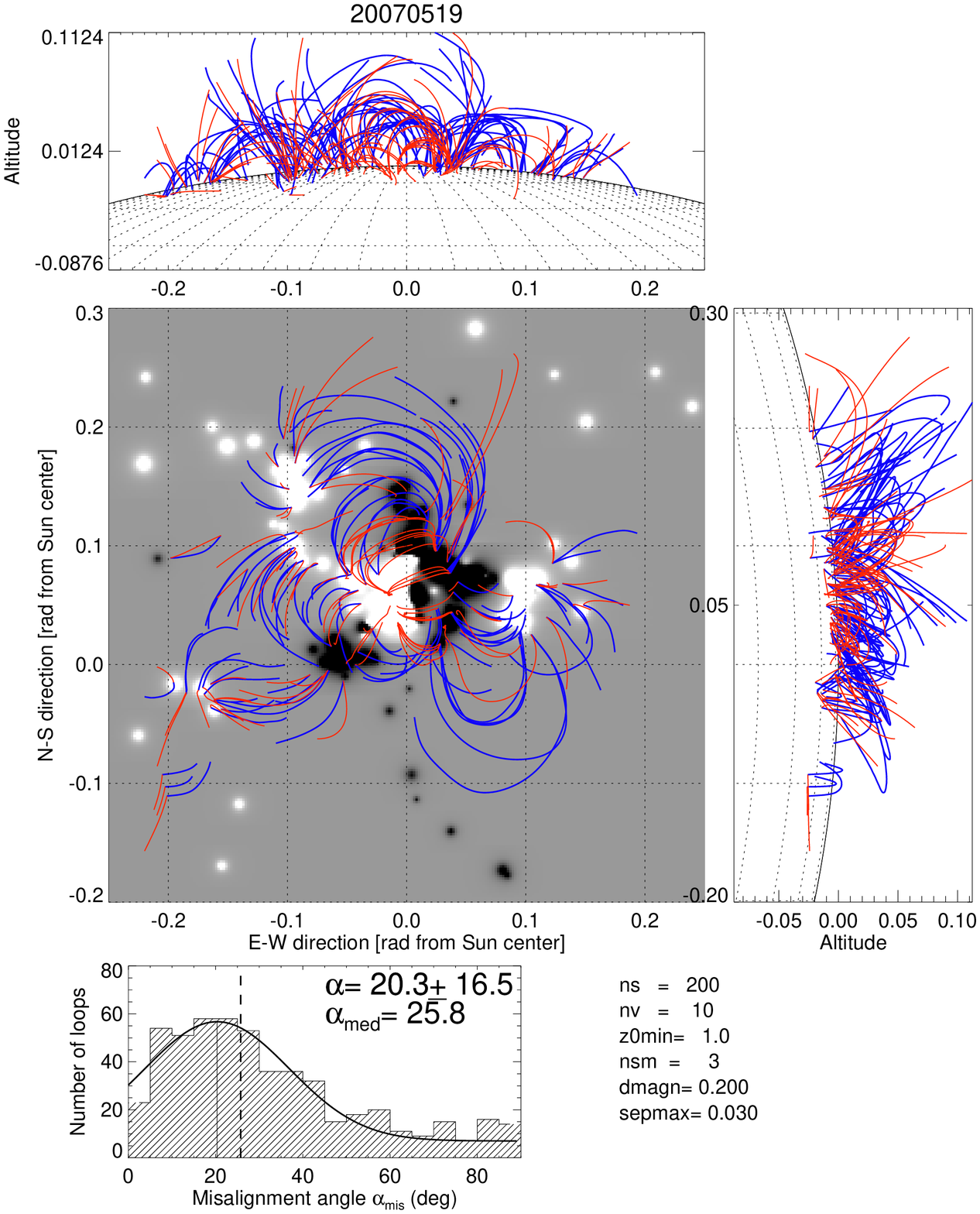}
\caption{Best-fit potential field model of AR observed on 2007 May 19.
Otherwise similar representation as Fig.~5.}
\end{figure}

\begin{figure} 
\plotone{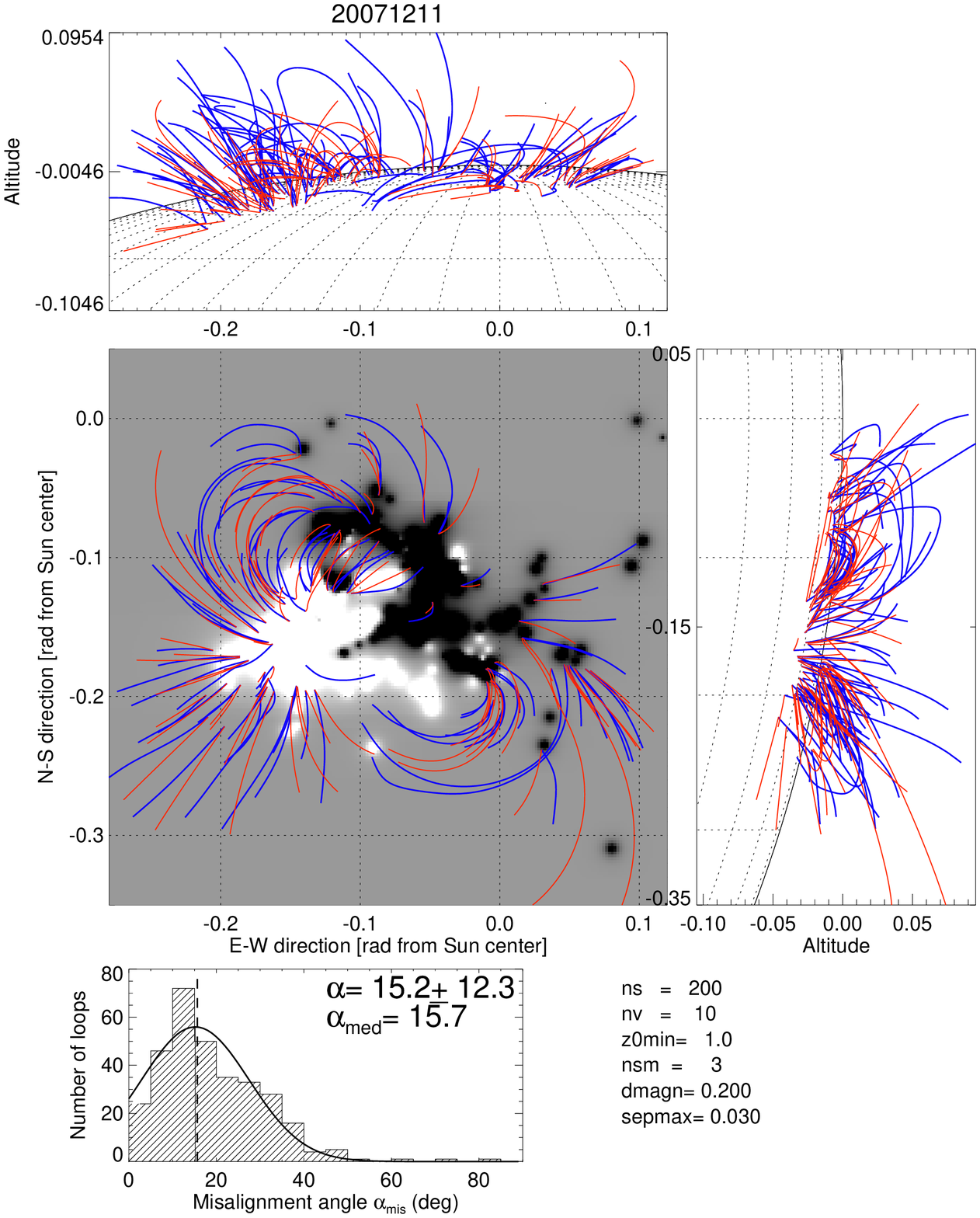}
\caption{Best-fit potential field model of AR observed on 2007 Dec 11.
Otherwise similar representation as Fig.~5.}
\end{figure}

\begin{figure} 
\plotone{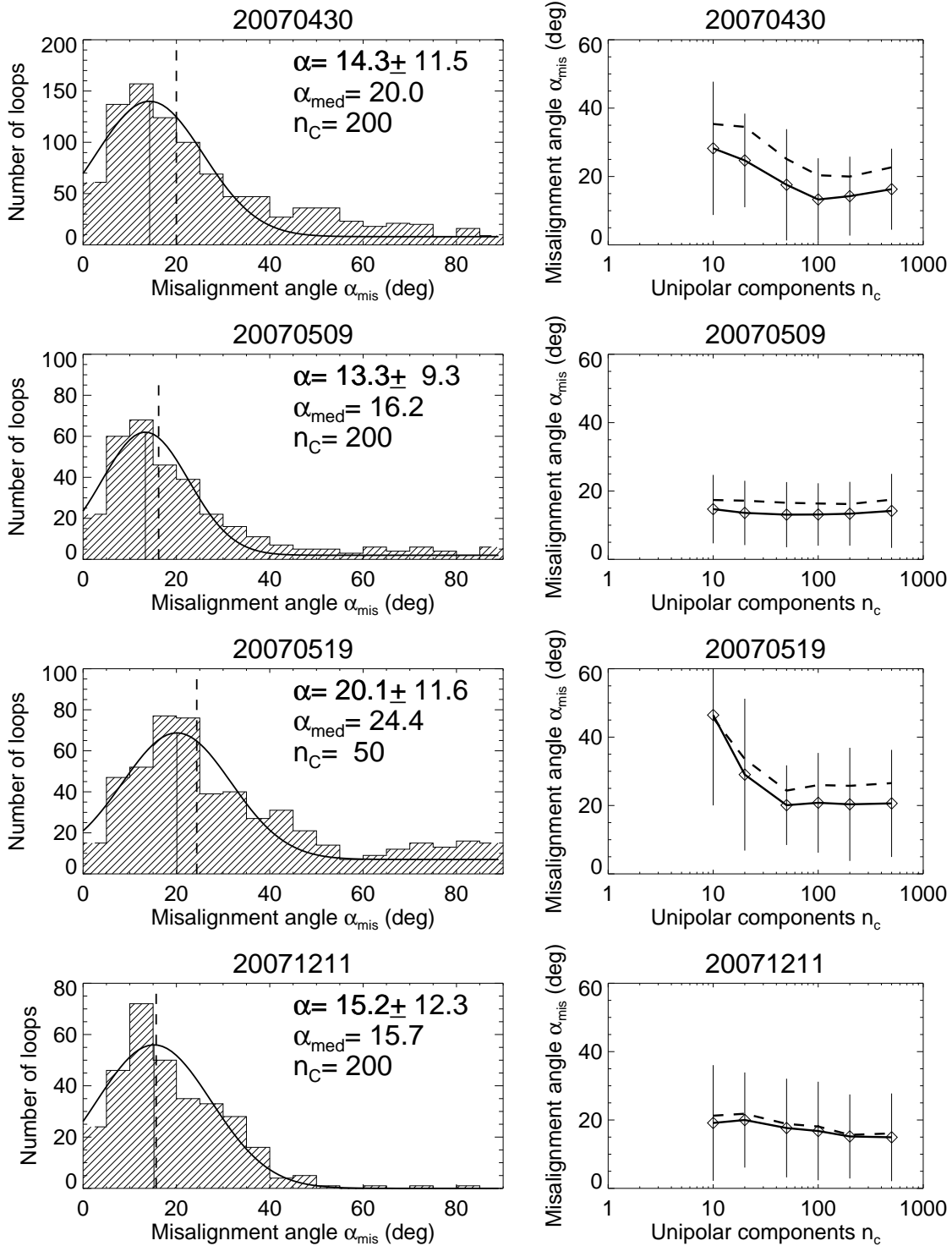}
\caption{Mean misalignment angle of best-fit potential field solution
as a function of the number $n_c$ of unipolar components (right side)
and distribution of misalignment angles for one particular value of $n_c$
(left side). The solid curves denote the mean of the misalignment angles,
and the dashed curves denote the median misalignment angles.}
\end{figure}

\begin{figure} 
\plotone{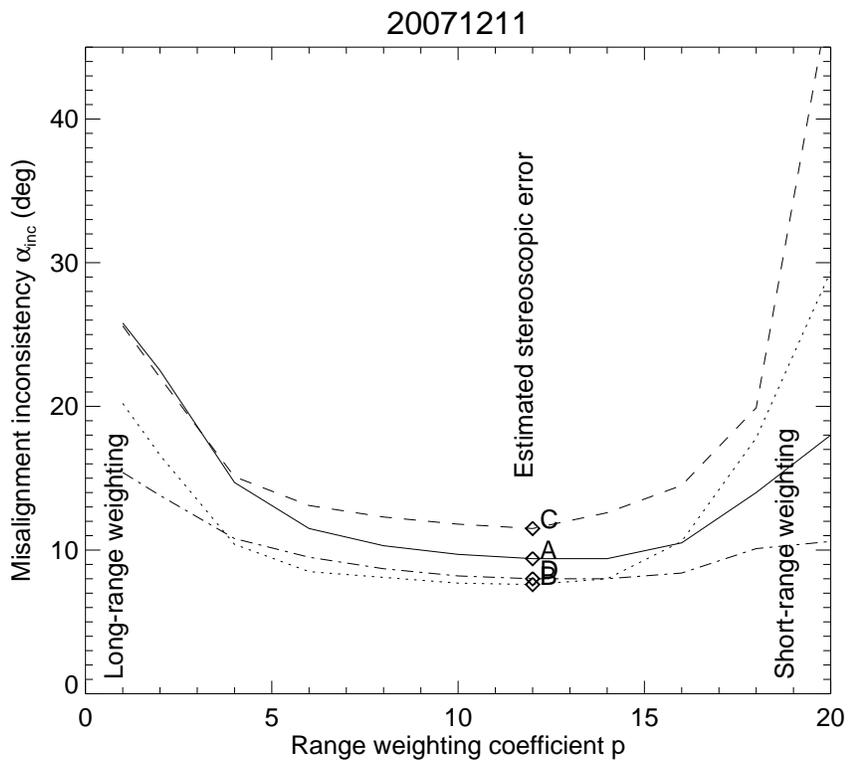}
\caption{Evaluation of stereoscopic error from the consistency of
misaligment angles as a function of the short- or long-range weighting.}
\end{figure}

\begin{figure} 
\plotone{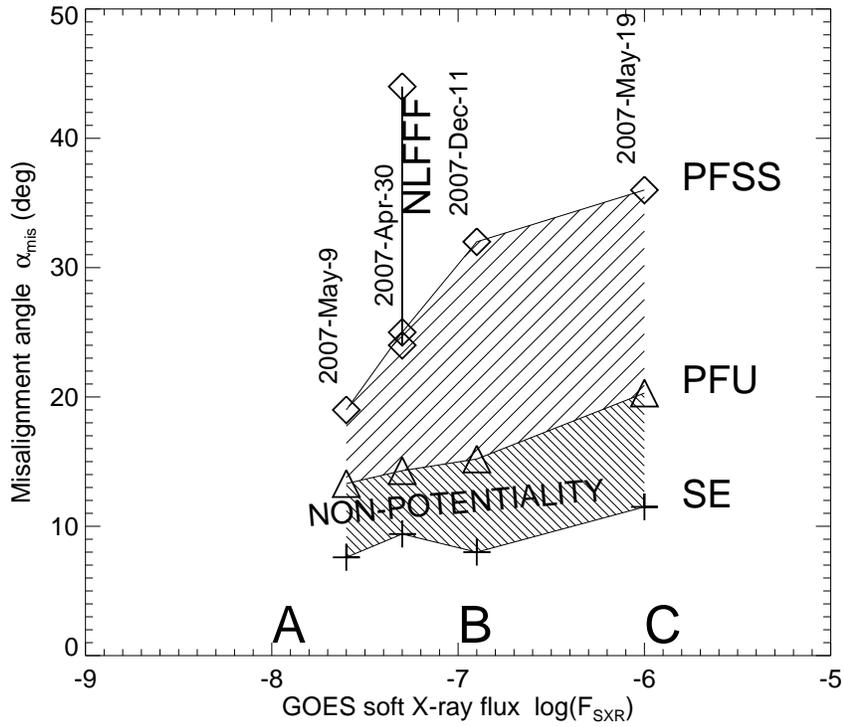}
\caption{The mean misalignment angle for four active regions as a function
of the GOES soft X-ray flux: for the potential field source surface model 
(PFSS: diamonds), 
for the unipolar potential field model bootstrapped with observed STEREO 
loops (PFU: triangles), and contributions from stereoscopic measuement errors 
(SE; crosses). The difference between the best-fit potential field model
(triangles) and stereoscopic errors (crosses) can be considered as a
measure of the non-potentiality of the active region (hatched area).}  
\end{figure}

\begin{figure} 
\plotone{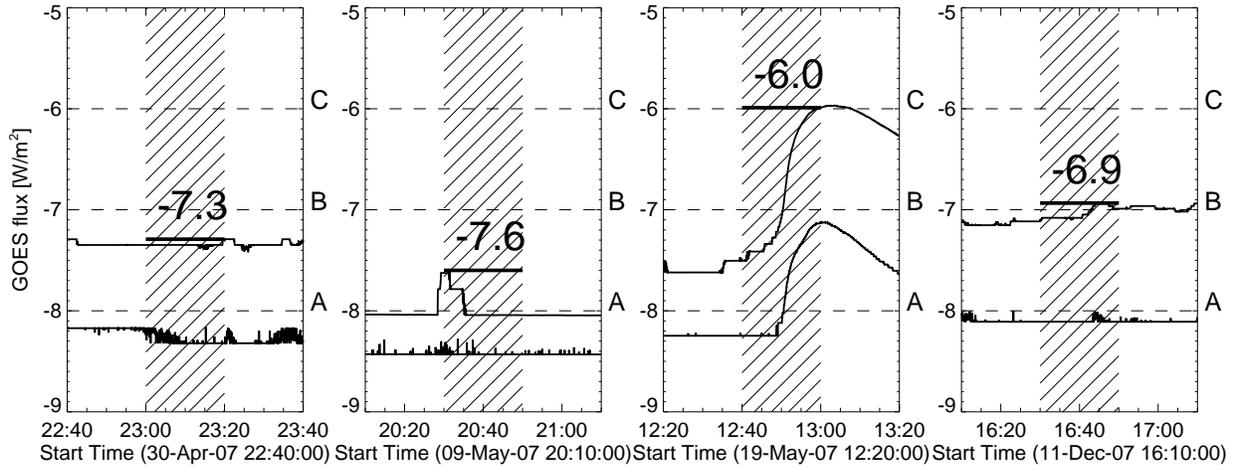}
\caption{GOES soft X-ray light curves of the 0.5-4 \ang\ (upper curve)
and 1-8 \ang\ channel (lower curve)
during the time of stereoscopic triangulation and magnetic modeling of the
active region. The peak level of the GOES flux during the observing time
is indicated with a thick bar.}
\end{figure}

\end{document}